\documentclass[aps]{revtex4}
\usepackage{amssymb,epsf}

\begin{document}
	
\title{Three dimensional dilatonic gravity's rainbow: exact solutions}
\author{S. H. Hendi$^{1,2}$\footnote{email address: hendi@shirazu.ac.ir}, B. Eslam Panah$^{1,2}$ \footnote{email address: behzad.eslampanah@gmail.com} and S. Panahiyan$^{1,3}$ \footnote{email address: sh.panahiyan@gmail.com}} \affiliation{$^1$Physics
Department and Biruni Observatory, College of Sciences, Shiraz University, Shiraz 71454, Iran\\ $^2$ Research Institute for Astronomy and Astrophysics of
Maragha (RIAAM), P.O. Box 55134-441 Maragha, Iran \\ $^{3}$ Physics Department, Shahid Beheshti University, Tehran 19839, Iran}	

\begin{abstract}%
Deep relations of dark energy scenario and string theory results
into dilaton gravity, on one hand, and the connection between
quantum gravity with gravity's rainbow, on the other hand,
motivate us to consider three dimensional dilatonic black hole
solutions in gravity's rainbow. We obtain two classes of the
solutions which are polynomial and logarithmic forms. We also
calculate conserved and thermodynamic quantities, and examine the
first law of thermodynamics for both classes. In addition, we
study thermal stability and show that one of the classes is
thermally stable while the other one is unstable.
\end{abstract}

\maketitle

\section{Introduction\label{Intro}}

In the context of AdS/CFT correspondence, the Hawking-Page phase
transition is often applied for investigating a
confinement-deconfinement phase transition. In other words, in
lower than a critical temperature ($T_{c}$), quarks are confined
to be grouped together in pairs or triples configurations, while
for $T>>T_{c}$ they are in a deconfined phase (quark-gluon
plasma). In the black hole language, a transition from unstable to
thermally stable solutions is the so-called Hawking-Page phase
transition. The Hawking-Page phase transition is one of the
primitive attempts for obtaining a quantum theory of gravity. In
general, black hole thermodynamics is a bridge between classical
viewpoint of general relativity with quantum gravity. Although we
could not already find a full description of quantum theory of
gravity, the attempts for reducing the gaps between the theory of
general relativity and quantum mechanics are still alive.
Gravity's rainbow is one of such attempts that arises from a deep
insight to remove the gap between classical theory of gravity and
its quantum nature. Recently, it has also been observed that
gravity's rainbow has some interesting consequences. Among them
one may recall black hole remnant \cite{Remnant}, information
paradox \cite{paradox} and nonsingular universe
\cite{Nonsingular1,Nonsingular2,Nonsingular3}.

On the other hand, one of the most common results of quantum
gravity is the existence of a minimum length
\cite{MinimumL-QG1,MinimumL-QG2}. Also, this minimum length may be
naturally arisen in string theory
\cite{MinimumL-String1,MinimumL-String2,MinimumL-String3}. The
existence of a minimum measurable length can be translated to an
upper limit of energy probe in high energy physics. In addition to the constant velocity of light, one may regard invariant Planck energy
to obtain doubly special relativity \cite{DSR1,DSR2,DSR3}. The
doubly special relativity is motivated by the following
generalized energy-momentum dispersion relation
\begin{equation}  \label{MDR}
E^2 f^2(\varepsilon)-p^2 g^2(\varepsilon)=m^2,
\end{equation}
where $\varepsilon=E/E_P$, $E_P$ is the Planck energy and $E$ is
the energy probe of the test particle. The functions
$f(\varepsilon)$ and $g(\varepsilon)$ are called rainbow
functions, and they should satisfy the following constraint
\begin{equation}
\lim\limits_{\varepsilon \to0} f(\varepsilon)=1,\qquad
\lim\limits_{\varepsilon \to0} g(\varepsilon)=1.
\end{equation}

This condition guarantees that one can reproduce the standard
dispersion relation in the infrared limit. Basically, it has also
been possible to extend doubly special relativity to the case of
curved spacetime and the resultant theory is known as doubly
general relativity or gravity's rainbow \cite{DGR1,DGR2}. The
arbitrary metric $\mathcal{G}(E)$ in gravity's rainbow can be
written as \cite{DGR2}
\begin{equation}  \label{rainmetric}
\mathcal{G}(E)=\eta^{ab}e_a(E)\otimes e_b(E),
\end{equation}
where the energy dependency of the frame fields are
\begin{eqnarray*}
&&e_0(E)=\frac{1}{f(\varepsilon)}\tilde{e}_0, \qquad e_i(E)=\frac{1}{g(\varepsilon)}\tilde{%
e}_i,
\end{eqnarray*}
in which the tilde quantities refer to the energy independent
frame fields as well. As we mentioned, the maximum attainable
energy is the Planck energy ($E_P$) and the energy at which
spacetime is probed is represented by $E$. Therefore, considering
a test particle which is probing the geometry of spacetime, one
can regard $E$ as the energy of that test particle.

On the other hand, it is important to study the UV completion of
general relativity such that it reduces to the general relativity
in the IR limit. In this regard, one may generalize the Einstein
theory to Horava-Lifshitz gravity \cite{HL1,HL2}. The
Horava-Lifshitz gravity is also motivated by a deformation of the
usual energy-momentum dispersion relation in the UV limit. So,
there is a close relation between gravity's rainbow and
Horava-Lifshitz gravity \cite{Garattini2015}. This relation is due
to the fact that the foundations of both theories are based on
breaking of the usual energy-momentum dispersion relation in the
UV limit. Considering the UV deformation of geometries that occurs
in string theory and the close relation between Horava-Lifshitz
gravity with gravity's rainbow \cite{Garattini2015}, one is
motivated to regard modified energy dependent metrics.

One of the main motivations of considering the string theory, at
present, is furnishing a combined quantum theory of gravity and
gauge field interactions \cite{StringQG1,StringQG2,StringQG3}. It
was shown that, in the context of string theory, the usual
gravitational field tensors are mixed with a scalar field partner;
the so-called dilaton. Taking into account the dilaton field, one
may ask for its coupling with other fields. It is considered that
dilaton may couple with gravitational field in Jordan frame
(Brans-Dicke theory), while it couples to matter field in Einstein
frame (dilaton gravity). Actions and field equations of both
Brans-Dicke theory and dilaton gravity are conformally related.
Therefore, one may use the solutions of Einstein frame to obtain
the corresponding ones in Jordan frame
\cite{Conformal1,Conformal2,Conformal3,Conformal4}.

The works of Fierz, Jordan, Brans and Dicke
\cite{Fierz1,Fierz2,Fierz3} were the pioneer attempts for coupling
the scalar field with gravity. Dilaton gravity theories were
renewed in the 1990ies for explanting the
observational results of supernovae at high values of the redshift \cite%
{Riess1,Riess2,Riess3,Riess4}, which are related to the evidences
for a nonvanishing dark energy \cite
{Wetterich1,Wetterich2,Wetterich3,Wetterich4,Wetterich5}.

Excessive interests of quantum gravity and dark energy scenarios
with their relations to dilaton gravity, motivate one to study
analytical solutions of Einstein gravity in the presence of a
dilaton field. In this paper, we obtain three dimensional black
hole solutions in the context of Einstein-dilaton theory with an
energy-dependent metric.

Three dimensional (charged) solutions, the so-called BTZ
(Banados---Teitelboim---Zanelli) black holes, have attracted a lot
of interest \cite{BTZ1,BTZ2,BTZ3,BTZ4,BTZ5,BTZ6,BTZ7}.
Generalization of BTZ solutions to nonlinear electrodynamics,
$F(R)$ theory and massive gravity have been investigated in Refs.
\cite{GenBTZ1,GenBTZ2,GenBTZ3,GenBTZ4,GenBTZ5,GenBTZ6,GenBTZ7,GenBTZ8massive}.
A class of higher dimensional modified nonlinear charged
solutions, BTZ-like black holes and wormholes, were obtained in
literature \cite{BTZlike1,BTZlike2}. Here, we are going to
generalize the solutions of \cite{PhysScripta} to the case of
energy dependent spacetime.

\section{Three dimensional dilatonic black holes in gravity's rainbow \label%
{Field}}

Regarding a minimal coupling between Einstein gravity and dilaton field ($%
\Phi $), in addition to the Ricci scalar ($\mathcal{R}$) (which is
related to Einstein gravity), one may regard a kinetic term
($\left( \nabla \Phi \right) ^{2}$) as well as a potential term
($V(\Phi )$). Thus the suitable
Lagrangian density of Einstein-dilaton gravity can be written as%
\begin{equation}
\mathcal{L}=\mathcal{R}-2\left( \nabla \Phi \right) ^{2}-V(\Phi ).
\label{Lagrangian}
\end{equation}

Using the variational method, we obtain the following equations of
motion
\begin{equation}
{R}_{\mu \nu }=2\partial _{\mu }\Phi \partial _{\nu }\Phi +g_{\mu
\nu }V(\Phi ),  \label{FE1}
\end{equation}%
\begin{equation}
\nabla ^{2}\Phi =\frac{1}{4}\frac{\partial V}{\partial \Phi },
\label{FE2}
\end{equation}%
in which we will use the following Liouville-type dilaton
potential
\begin{equation}
V(\Phi )=2\Lambda e^{2\alpha \Phi },  \label{v1}
\end{equation}%
where $\Lambda$ is a constant which is referred to the
cosmological constant and $\alpha$ is the dilaton parameter. Now,
we consider the following three dimensional energy-dependent
static metric
\begin{equation}
ds^{2}=-\frac{\Psi (r)}{f^{2}(\varepsilon )}dt^{2}+\frac{dr^{2}}{%
g^{2}(\varepsilon )\Psi
(r)}+\frac{r^{2}R^{2}(r)}{g^{2}(\varepsilon )}d\phi ^{2},
\label{Metric1}
\end{equation}%
where $f^{2}(\varepsilon )$ and $g^{2}(\varepsilon )$ are rainbow
functions.

Taking into account the field equations with the metric
(\ref{Metric1}), we find that $tt$, $rr$ and $\phi \phi $
components (the nonzero components) of Eq. (\ref{FE1}) can be,
respectively, simplified as
\begin{eqnarray}
e_{tt} &=&\Psi ^{\prime \prime }(r)+\frac{\Psi ^{\prime
}(r)}{r}+\frac{\Psi ^{\prime }(r)R^{\prime
}(r)}{R(r)}+\frac{2V(\Phi )}{g^{2}(\varepsilon )}=0,
\label{ett} \\
e_{rr} &=&e_{tt}+2\Psi (r)\left( \frac{2R^{\prime }(r)}{rR(r)}+\frac{%
R^{\prime \prime }(r)}{R(r)}+2\Phi ^{\prime 2}(r)\right) =0,  \label{err} \\
e_{\phi \phi } &=&\frac{\Psi ^{\prime }(r)}{r}+\frac{\Psi ^{\prime
}(r)R^{\prime }(r)}{R(r)}+\frac{2\Psi (r)R^{\prime
}(r)}{rR(r)}+\frac{\Psi (r)R^{\prime \prime
}(r)}{R(r)}+\frac{V(\Phi )}{g^{2}(\varepsilon )}=0. \label{epp}
\end{eqnarray}%

Besides, Eq. (\ref{FE2}) may be rewritten as
\begin{equation}
4\Psi ^{\prime }(r)\Phi ^{\prime }(r)+4\Psi (r)\left( \Phi
^{\prime \prime
}(r)+\frac{\Phi ^{\prime }(r)}{r}+\frac{R^{\prime }(r)\Phi ^{\prime }(r)}{%
R(r)}\right) -\frac{1}{g^{2}(\varepsilon )}\frac{dV(\Phi )}{d\Phi
}=0. \label{d2phi}
\end{equation}%

Considering both Eqs. (\ref{ett}) and (\ref{err}), simultaneously
(regarding $e_{tt}-e_{rr}$), one can find that the second part of
Eq. (\ref{err}) should vanish, separately
\begin{equation}
e_{tt}-e_{rr}=\frac{2R^{\prime }(r)}{rR(r)}+\frac{R^{\prime \prime }(r)}{R(r)}%
+2\Phi ^{\prime 2}(r)=0,  \label{err-ett}
\end{equation}%
which is independent of the metric function ($\Psi(r)$). After
some manipulations, we find
\begin{equation}
e_{tt}-e_{rr}=\frac{2}{r}\frac{d}{dr}\ln R(r)+\frac{d^{2}}{dr^{2}}\ln
R(r)+\left[ \frac{d}{dr}\ln R(r)\right] ^{2}+2\Phi ^{\prime
2}(r)=0.  \label{err-ett2}
\end{equation}

Taking into account Eq. (\ref{err-ett2}), one finds that $R(r)$ is
an exponential function of $\Phi(r)$. In what follows, we regard
two classes of $R(r)$ function to obtain exact solutions.

\subsection{\textbf{Case I: $R(r)=e^{\protect\beta \Phi }$ with }$\protect%
\beta $\textbf{$\neq 2$:}}

Now, we regard the ansatz $R(r)=e^{\beta \Phi }$ to obtain the
unknown functions. Using the field equations with the line element
(\ref{Metric1}), we obtain
\begin{equation}
\Phi (r)=\sqrt{\gamma \left( 1-2\gamma \right) }\ln \left( \frac{b}{r}%
\right) ,  \label{phi}
\end{equation}%
\begin{equation}
\Psi (r)=\frac{-\Lambda r^{2}}{\gamma (6\gamma
-1)g^{2}(\varepsilon )}\left( \frac{b}{r}\right) ^{2(1-2\gamma
)}-\frac{m}{r^{2\gamma -1}},  \label{f(r)}
\end{equation}%
where $b$ and $m$ are integration constants, $\gamma =1/(2+\beta
^{2})$ (with constraint $\gamma<\frac{1}{2}$) and $\Lambda
=-l^{-2}$. Inserting Eqs. (\ref{phi}) and (\ref{f(r)}) in the
field equations (\ref{FE1}) and (\ref{FE2}) with arbitrary $\beta
\neq 2$ ($\gamma
\neq 1/6$), one can show that all the equations are satisfied only for $%
\beta =\alpha $.

Looking for the black hole solutions, one should study the
curvature scalars. Calculations show that the Kretschmann scalar
diverges at the origin ($r=0$) and is finite for $r>0$. Thus one
concludes that there is an essential singularity located at $r=0$.
This singularity can be covered with an event horizon, and
therefore, we can interpret the solution as black hole. The root
of the metric function is located at
\begin{equation}
r_{+}=\left( \frac{b^{2(1-2\gamma )}}{\gamma (6\gamma
-1)l^{2}mg^{2}(\varepsilon )}\right) ^{\frac{1 }{ \left( 1-6\gamma
\right) }}.  \label{Horizon}
\end{equation}

Eq. (\ref{Horizon}) shows that in order to have a real horizon
radius with positive mass and rainbow function, we should regard
$1/6<\gamma <1/2$.

\begin{figure}[tbp]
	$%
	\begin{array}{c}
	\epsfxsize=5cm \epsffile{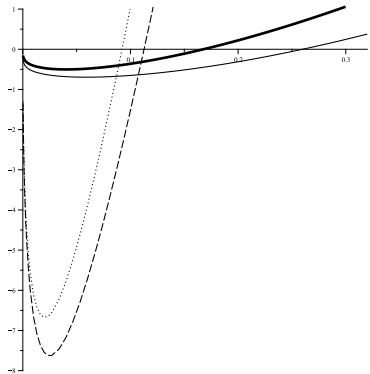}
	\end{array}
	$%
	\caption{The
		behavior of $\Psi(r)$ versus $r$ for $b=2$, $l=0.5$, $m=1.4$ and
		$g(\protect\varepsilon)=1.01$.\newline
		Eq. (\protect\ref{f(r)}): $\protect\alpha =0.5$ (solid line) and $\protect%
		\alpha =0.7$ (bold line).\newline Eq. (\protect\ref{f(r)2}):
		$L=0.09$ (dotted line) and $L=0.11$ (dashed line).} \label{Met1}
\end{figure}

\begin{figure}[tbp]
	$%
	\begin{array}{c}
	\epsfxsize=5cm \epsffile{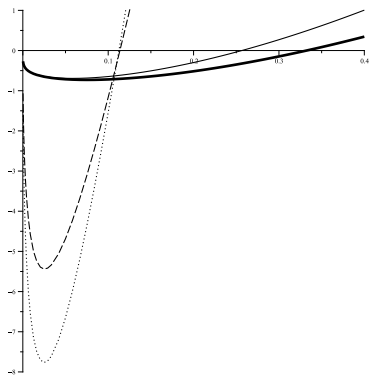}
	\end{array}
	$%
\caption{The behavior of $\Psi(r)$ versus $r$ for $b=2$, $l=0.5$ and $m=1.4$.%
	\newline
	Eq. (\protect\ref{f(r)}): $\protect\alpha =0.5$, $g(\protect\varepsilon%
	)=1.001$ (solid line) and $g(\protect\varepsilon)=1.2$ (bold
	line).\newline Eq. (\protect\ref{f(r)2}): $L=0.11$,
	$g(\protect\varepsilon)=1.001$ (dotted line) and
	$g(\protect\varepsilon)=1.2$ (dashed line).} \label{Met2}
\end{figure}

It is notable that in the absence of dilaton field
($\alpha=\beta=0$ and $\gamma =1/2$), obtained solution reduces to
the BTZ black hole with AdS asymptote. In order to investigate the
effects of dilaton field, we plot the metric function. Regarding
Figs. \ref{Met1} and \ref{Met2}, one finds that although the
singularity is spacelike in the absence of dilaton field,
dilatonic solutions contain a null-like singularity at the origin.
In other words, metric function has a real positive root for
$\alpha =0$, while there are two real nonnegative roots for
dilatonic solution, in which inner horizon is located at $r=0$
($\lim_{r \rightarrow 0^{+}} \Psi(r)=0$). Furthermore, one can
find that although rainbow functions may affect the location of
the event horizon, they do not change the type of singularity (see
Fig. \ref{Met2}). In addition, dilaton field affects the
asymptotical behavior and obtained solution is not asymptotically
AdS for $\alpha \neq 0$.

\subsection{\textbf{Case II: $R(r)=e^{\protect\beta \Phi }$ with
}$\protect\beta $\textbf{$=2$:} \label{alphatwo}}

In this section, we consider a special case, which was ill-defined
in previous solutions (Eq. (\ref{f(r)}). Regarding $\beta =2$, one
can find new field equations. It is straightforward to show that
Eqs. (\ref{FE1}) and (\ref{FE2}) have the following solutions for
$\beta =2$ ($R(r)=e^{2\Phi }$)
\begin{equation}
\Phi (r)=\frac{1}{3}\ln \left( \frac{b}{r}\right) ,  \label{phi2}
\end{equation}%
\begin{equation}
\Psi (r)=\left[ \frac{-6\Lambda b^{4/3}}{g^{2}(\varepsilon )}\ln (\frac{r}{L}%
)-m\right] r^{2/3},  \label{f(r)2}
\end{equation}%
where $m$ is an integration constant and $L$ is an arbitrary
constant with
length dimension. In order to have consistent solutions, we should set $%
\alpha =\beta =2$. Calculation of scalar curvatures shows that
there is a null singularity at the origin. Such as previous case,
this singularity can be covered by an event horizon with the
following real positive radius
\begin{equation}
r_{+}=L\exp \left( \frac{ml^{2}g^{2}(\varepsilon )}{6
b^{4/3}}\right) , \label{Horizon2}
\end{equation}%
where confirms that the event horizon of this black hole is
sensitive to variations of $L$ and rainbow functions (see Figs.
\ref{Met1} and \ref{Met2} for more details). Regarding the
asymptotical behavior of the
metric function for large $r$, we find that although $\Psi (r)\propto $ $%
\Lambda r^{2/3}\ln r$ confirms that\ the dominant term of Eq.
(\ref{f(r)2}) is $\Lambda -$term, it differs from the behavior of
asymptotically AdS spacetime, in which $\Psi (r)\propto $ $\Lambda
r^{2}$ for large $r$.\bigskip

\section{Thermodynamics of the solutions \label{Therm}}

In this section, we are going to obtain thermodynamic and
conserved quantities related to the solutions. At first, we
calculate the Hawking temperature. Using the surface gravity
interpretation, one finds that
\begin{equation}
T_{H}=\left. \frac{g(\varepsilon )}{f(\varepsilon )}\frac{\Psi ^{\prime }(r)%
}{4\pi }\right\vert _{r=r_{+}}.  \label{T1}
\end{equation}

After some simplifications, we obtain
\begin{equation}
T_{H}=\frac{1}{2\pi l^{2}f(\varepsilon )g(\varepsilon )}\left\{
\begin{array}{cc}
\frac{b^{2(1-2\gamma )}}{2\gamma }r_{+}^{4\gamma -1}, & \alpha \neq 2 \\
\frac{3b^{4/3}}{r_{+}^{1/3}}, & \alpha =2%
\end{array}%
\right. ,  \label{Tem}
\end{equation}%
where $r_{+}$ was obtained for each branches (see Eqs. (\ref{Horizon}) and (%
\ref{Horizon2})). Regarding positive rainbow functions, one
obtains positive definite temperature.

Now, we should obtain finite mass. AdS/CFT correspondence
\cite{Mal1,Mal2} guarantees that we can apply the counterterm
method for asymptotically AdS solutions to calculate finite
conserved quantities. Although, obtained
solutions are not asymptotically AdS, it was shown that \cite%
{flathorizon1,flathorizon2} one can find an appropriate
counterterm for removing the divergences in horizon-flat
spacetimes. The action of three-dimensional Einstein-dilaton
gravity with suitable Gibbons-Hawking surface term and counterterm
action can be written as
\begin{equation}
I_{total}=I_{bulk}+I_{boundary}+I_{counterterm},  \label{Act}
\end{equation}
in which
\begin{eqnarray}
&&I_{bulk} =-\frac{1}{16\pi
}\int_{\mathcal{M}}d^{3}x\sqrt{-g}\left(
R-2\left( \nabla \Phi \right) ^{2}-V(\Phi )\right)  \nonumber \\
&&I_{boundary}=-\frac{1}{8\pi }\int_{\partial \mathcal{M}}d^{2}x\sqrt{-h}%
\Theta (h),  \nonumber \\
&&I_{counterterm}= -\frac{1}{8\pi }\int_{\partial \mathcal{M}}d^{2}x\sqrt{-h}%
\left( \frac{1}{L_{\mathrm{eff}}}\right),
\end{eqnarray}
where $\Theta $ represents the trace of the extrinsic curvature of
the boundary ${\partial \mathcal{M}}$ with induced metric $h^{ab}$
and suitable effective length is
\begin{equation}
L_{\mathrm{eff}}=\sqrt{\frac{2\alpha ^{2}-2}{V(\Phi )}},
\label{L}
\end{equation}
which reduces to $l$ in the absence of the dilaton field ($\alpha
=0$), as expected. Considering the total finite action, we apply
the Brown-York method for constructing a divergence free
stress-energy tensor \cite{BY}
\begin{equation}
T^{ab}=\frac{1}{8\pi }\left( \Theta ^{ab}-\Theta h^{ab}-\frac{h^{ab}}{L_{%
\mathrm{eff}}}\right).  \label{Stres}
\end{equation}

Taking into account temporal Killing vector $\partial /\partial
t$, one can find its associated conserved quantity, which is the
quasilocal mass
\begin{equation}
{M}=\frac{m}{4f(\varepsilon )}\times \left\{
\begin{array}{cc}
\gamma b^{1-2\gamma }, & \alpha \neq 2 \\
\frac{b^{2/3}}{6}, & \alpha =2%
\end{array}%
\right. .  \label{M}
\end{equation}

It is known that the area law is valid in Einstein-dilaton
gravity. In addition, the validity of the area law is examined in
spacetimes with zero-curvature boundary of $t=cte$ and $r=cte$ in
the context of non-Einsteinian gravity. Therefore, we regard the
area law for our solutions to obtain entropy as
\begin{equation}
{S}=\frac{\pi }{2g(\varepsilon )}\times \left\{
\begin{array}{cc}
b^{1-2\gamma }r_{+}^{2\gamma }, & \alpha \neq 2 \\
b^{2/3}r_{+}^{1/3}, & \alpha =2%
\end{array}%
\right. .  \label{ent}
\end{equation}

Here, we are going to check the validity of the first law of
thermodynamics. To do this, we can obtain the mass ($M$) as a
function of the only extensive quantity ($S$). Regarding the fact
that metric function should vanish at the event horizon
($f(r_{+})=0$), we obtain the following Smarr-type formula
\begin{equation}
M(S)=\frac{-\Lambda b^{2}}{4f(\varepsilon )g^{2}(\varepsilon
)}\times \left\{
\begin{array}{cc}
\frac{1}{(6\gamma -1)}\left( \frac{b}{r_{+}}\right) ^{1-6\gamma },
& \alpha
\neq 2 \\
\ln (\frac{r_{+}}{L}), & \alpha =2%
\end{array}%
\right. ,  \label{Msmarr}
\end{equation}%
where%
\begin{equation}
r_{+}=\left\{
\begin{array}{cc}
\left( \frac{2g(\varepsilon ){S}}{\pi b^{1-2\gamma }}\right)
^{1/2\gamma },
& \alpha \neq 2 \\
\left( \frac{2g(\varepsilon ){S}}{\pi b^{2/3}}\right) ^{3}, & \alpha =2%
\end{array}%
\right. .
\end{equation}

Now, we can calculate the Hawking temperature as the intensive
quantity conjugate to $S$
\begin{equation}
T_{H}=\frac{dM}{dS}=\left( \frac{\partial M}{\partial
r_{+}}\right) \left( \frac{\partial S}{\partial r_{+}}\right)
^{-1},  \label{T2}
\end{equation}%
which is in agreement with Eq. (\ref{Tem}), and therefore, the
first law of thermodynamics is valid as
\begin{equation}
dM=T_{H}dS.  \label{1stLaw}
\end{equation}

\section{Thermodynamic stability \label{STAB}}

It is well-known that asymptotically flat uncharged black holes
(Schwarzschild solutions) are thermally unstable, and therefore,
in order to obtain stable black holes, one can insert cosmological
constant or electric charge to the solutions. Here, we are going
to study the effects of dilaton field on thermal stability of
neutral black holes.

Thermal stability of a system can be discussed in various
ensembles. In the canonical ensemble, thermal stability of a
system is determined by its heat capacity. The positivity of the
heat capacity ($C=T(\partial S/\partial T)$) is sufficient to
ensure thermal stability of a thermodynamical system. Regarding
obtained results for the temperature and entropy, one finds
\begin{equation}
C=\frac{T_{H}}{\frac{\partial T_{H}}{\partial S}}=\left\{
\begin{array}{cc}
\frac{\pi \gamma b^{1-2\gamma }r_{+}^{2\gamma }}{g(\varepsilon )(4\gamma -1)}%
, & \alpha \neq 2 \\
-\frac{\pi b^{2/3}r_{+}^{1/3}}{2g(\varepsilon )}, & \alpha =2%
\end{array}%
\right. .  \label{HeatCap}
\end{equation}

\begin{figure}[tbp]
	$%
	\begin{array}{c}
	\epsfxsize=5cm \epsffile{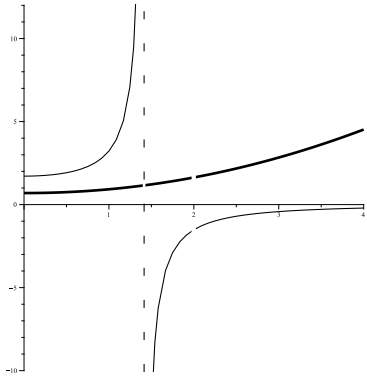}
	\end{array}
	$%
\caption{\textbf{Case $\protect\alpha \neq 2$:} The heat capacity
	(solid line) and
	temperature (bold line) versus $\protect\alpha$ for $b=1$, $l=0.5$, $f(%
	\protect\varepsilon)=1$, $g(\protect\varepsilon)=1.1$, and
	$r_{+}=1.2$.} \label{C}
\end{figure}

According to Eq. (\ref{HeatCap}), we find that the solutions are
unstable for $\alpha =2$ and $\left\vert \alpha \right\vert
>\sqrt{2}$. In other words, although BTZ black hole is stable
everywhere, three dimensional
uncharged dilatonic black holes are stable for $\alpha \in \left( -\sqrt{2},+%
\sqrt{2}\right) $ (or equivalently $1/4<\gamma <1/2$). As a
result, one may
regard a phase transition for the critical value of the dilaton parameter ($%
\alpha _{c}=\sqrt{2}$) (see Fig. \ref{C}, in which shows an
increasing behavior for positive $T$ and a divergence point for
the heat capacity).

\section{Conclusion} \label{CON}

In this paper, we have considered three dimensional gravity's
rainbow in the presence of a dilaton field. In order to obtain
consistent solutions, we have regarded a class of Liouville-type
dilaton potential. We have obtained two classes of the solutions
with polynomial and logarithmic forms. We also showed that the
solutions can be considered as a BTZ dilatonic black hole only for
$\alpha^{2}\leq 4$.

Taking into account the area law and the surface gravity
interpretation, we have calculated the entropy and the temperature
of the black hole solutions. We have also added a suitable
counterterm to the gravitational action to obtain finite mass. In
addition, using the Smarr-type formula for the mass, we have found
that the first law of thermodynamics is valid for both classes of
the solutions.

Then, we have investigated thermal stability of the solutions and
calculated the heat capacity. We have found that logarithmic black
hole solution is unstable while polynomial one may be thermally
stable.

Finally, it is worthwhile to investigate stationary and/or charged
solutions as well as their extension to massive gravity
\cite{Massive1,Massive2}. In addition, it is interesting to study
thermodynamic geometry of the solutions based on HPEM (Hendi, Panahiyan, Eslam Panah and Momennia) method
\cite{HPEM1,HPEM2}. We will address these issues in the
forthcoming work.

\section*{Acknowledgment}

We gratefully thank the anonymous referee for enlightening
comments and suggestions which substantially helped in proving the
quality of the paper. We also wish to thank the Shiraz University
Research Council. This work has been supported financially by
Research Institute for Astronomy and Astrophysics of Maragha
(RIAAM), Iran.


%

\end{document}